\documentclass[12pt,preprint]{aastex}

\begin{document}

\title{The Location of the Nucleus and the Morphology of Emission Line
Regions in NGC 1068}

\author{Rodger I. Thompson}
\affil{Steward Observatory, University of Arizona,
    Tucson, AZ 85721}
\email{rthompson@as.arizona.edu}

\author{Ranga-Ram Chary}
\affil{Lick Observatory, University of California,
    Santa Cruz, CA 95064}
\email{rchary@ucolick.org}

\author{Michael R. Corbin}
\affil{Steward Observatory, University of Arizona,
    Tucson, AZ 85721}
\email{mcorbin@as.arizona.edu}

\author{Harland Epps}
\affil{Lick Observatory, University of California,
    Santa Cruz, CA 95064}
\email{epps@ucolick.org}

\begin{abstract}

This paper presents new NICMOS data on the location of the nucleus and
the morphology of hydrogen and [SiVI] emission in NGC 1068. The peak of
the emission at 2.2 $\micron$ is a strong point source which marks the
location of the nucleus. The [SiVI] line emission region consists of
two main components, a diffuse region of coronal emission to the
north-northeast of the nucleus and a bright emission spot $1.6 \arcsec$
from the nucleus along the direction of the radio jet.  A similar but
less intense emission spot also occurs in the hydrogen Paschen
$\alpha$  and WFPC2 H$\alpha$ images.  The accurate determination of
the nuclear position and its relation to the emission line morphology
produces a clearer picture of the nature of the interaction between the
radio jet and its surroundings.

\end{abstract}


\section{INTRODUCTION}

As the nearest Seyfert 2 galaxy, NGC 1068 is the subject of many
morphological studies.  Since the central engine is hidden from direct
optical view by surrounding dust, the exact location of the nucleus has
been the subject of some debate.  Previous optical studies with the
Hubble Space Telescope (HST) \citep{cap97,cap95} inferred the nuclear
position  from polarization and astrometry studies.  Ground based
studies at wavelengths longer than 2 $\micron$ directly image the
nucleus \citep{bk00,bk98,that97} but are limited by atmospheric seeing
or the accuracy of the astrometry relative to galactic features. The
NICMOS camera 2 observations directly observe the nucleus at 2.2
$\micron$ and image known emission line features in the Paschen
$\alpha$ line, firmly establishing the location of the nucleus relative
to those features.

The existence of coronal [SiVI], [SiVII] and [SiX] emission in NGC 1068 is
well established by ground based spectroscopic studies
\citep{Om90,Mo93,thm96}. The relative strengths of the silicon emission
lines pointed to a hard spectrum photoionizing source as the primary
excitation for the majority of the emission. It was assumed that this
source was the nuclear central engine.  The [SiVI] image presented in
this work shows that the [SiVI] emission is spatially separated into
at least two components with the possibility of different excitation
sources for each component as discussed in Section~\ref{int}.  A
similar separation of excitation components has been suggested by
\citet{krm00}.

\section{OBSERVATIONS}

The observations utilized in this work are part of an extensive imaging
campaign with the HST infrared instrument, NICMOS, on two classic
Seyfert galaxies, NGC 1068 and NGC 4151.  This study utilizes the line
and continuum filters for the [Fe II] 1.64 $\micron$ line, (F164N,
F166N), Paschen $\alpha$ 1.875 $\micron$ line, (F187N, F190N) and [SiVI] 
1.96 $\micron$ line, (F196N, F200N).   The camera 2 molecular
hydrogen and continuum filter images were also inspected to insure that
the [SiVI] images were not contaminated with molecular hydrogen
emission from the H$_{2}$ S5 (1-0) line emission at 1.97 $\micron$.
The small redshift of NGC 1068 (z = 0.0038) shifts part of the line
emission out of the bandpass of the narrow band line filters so
quantitative measurements are difficult. From ground based infrared
spectra and the width of the filters we estimate that about 1/3 of the
flux falls outside of the bandpass, depending on the intrinsic width of
the line.  All of the images were taken in an eight point spiral dither
pattern.  The dither step sizes are $0.27 \arcsec$ for camera 1 and 2
and $0.54 \arcsec$ for camera 3 observations.  The WFPC2 F658N filter
minus continuum H$\alpha$ image is from the HST archive (proposal 5754)
and has been processed to remove cosmic ray hits and other features. It
is discussed in \citet{cap97b}.

\section{DATA REDUCTION} 
The NICMOS images were reduced utilizing IDL based procedures developed
for image production in the Hubble Deep Field \citep{thm99}. The strong
point-like emission from the nucleus was used to accurately align the
dithered images.  Emission line images were produced by subtraction of
the associated continuum filter images from the line filter images.
Although the line and continuum filter wavelengths are very close
together, there is still a contiuum level difference between the line
and continuum filters in very red objects such as the nucleus of NGC
1068.  To compensate, the intensity of the continuum image was scaled
to produce the best subtraction measured by the strength of the
residuals in the nuclear region.  Even with this adjustment the high
contrast between the nucleus and the surrounding line emission produces
significant residuals within the nuclear region.  The area contained
within the first two Airy rings does not accurately reflect the true
emission levels and is not used in this analysis.  High residuals are
not unexpected since the ratio of nuclear continuum emission to line
emission is 125 to 1 for Paschen $\alpha$ and 4000 to 1 for [SiVI].

\section{POSITION OF THE NUCLEUS} \label{sec-nuc}

In all NICMOS images at wavelengths of 1.6 $\micron$ or longer there is
a strong, unresolved, extremely red, source that displays Airy rings
and a diffraction pattern.  We assume that this source is either the
nucleus or the emission from hot dust centered on the nucleus.
\citet{thmc99} put a limit on the width of the point source as $0.03
\arcsec$ which corresponds to a diameter of 2 pc in NGC 1068 based on
the distance of 14 Mpc for a Hubble constant of 75 km sec$^{-1}$
Mpc$^{-1}$.  This is consistent with the $0.02 \arcsec$ limit of
\citet{wein99} utilizing ground based speckle observations. The point
source accurately marks the position of the nucleus relative to the
emission features observed in the line images.  Recent radio
observations, \citet{roy98} e.g., have accurately mapped the radio
emission in the region and identified the source designated S1 as the
position of the nucleus.  VLBA observations by \citet{gal97} of a disk
with H$_2$O and OH maser emission from the S1 source leaves little
doubt that this is the nuclear position.  We assume that S1 and our
observed nuclear point source are coincident in the following analysis.
The very similar features between the NICMOS Paschen $\alpha$ and the
WFPC H$\alpha$ image then register the nuclear position and the radio
emission to both the optical and infrared emission line features.
Within the error bars, the nuclear position determined from the NICMOS
images is consistent with the position given by \citet{kis99} in his
reanalysis of the FOC data used by \citet{cap95}.  It is also
coincident with the optical feature designated cloud B by
\citet{ev91}.  The coincidence is accurate to 0.04 arc seconds, the
half width of a NICMOS camera 2 pixel.

\section{MORPHOLOGY OF THE EMISSION REGION}

As is the case with the optical emission, each stretch of the image
emphasizes a particular aspect of the emission morphology.  No single
stretch reveals all of the features so we will describe some of the
features in the text.

\subsection{Hydrogen Emission Region}

\begin{figure}

\scalebox{.7}{\includegraphics{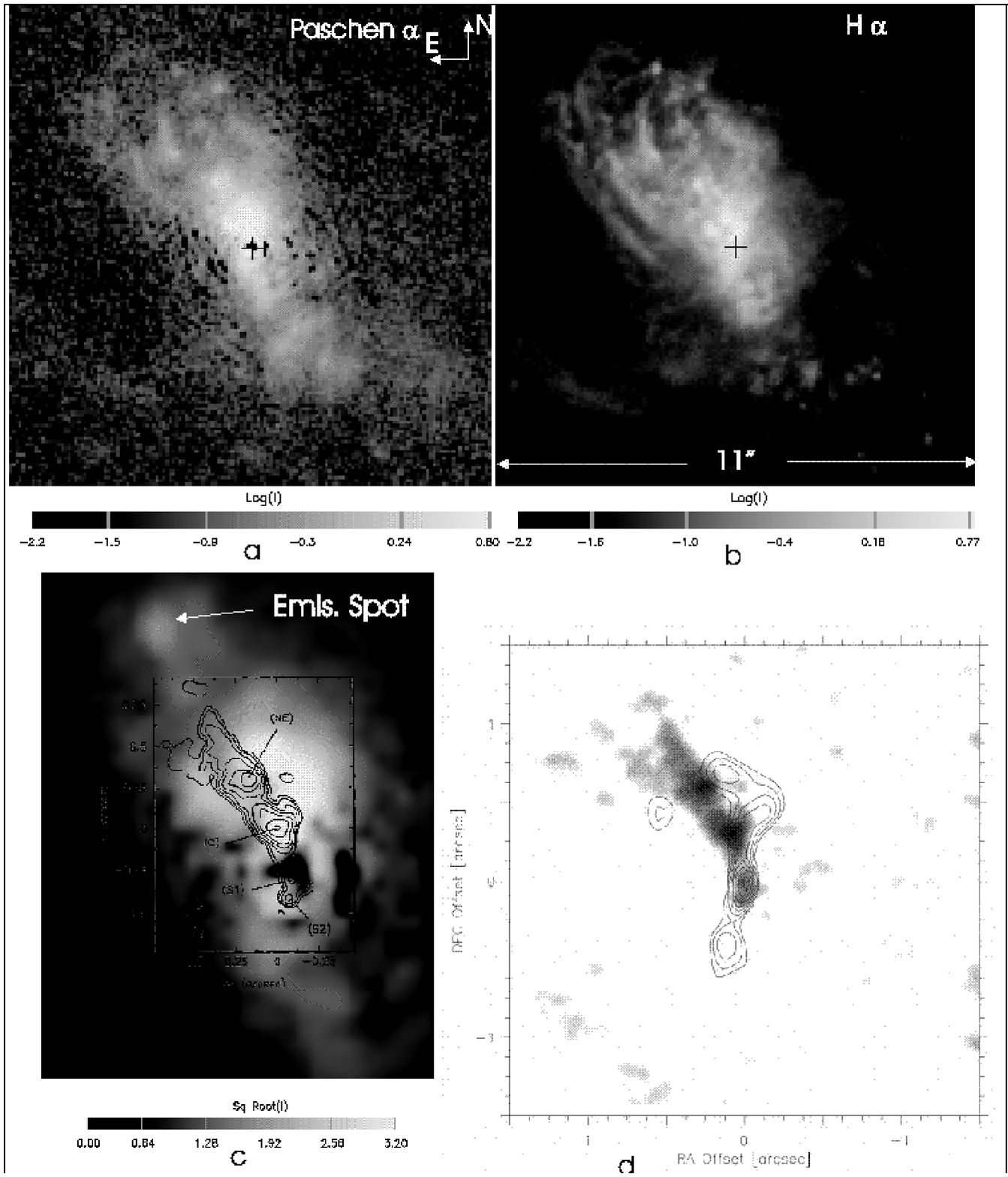}}

\caption{Fig. 1a and b are the Paschen $\alpha$ and H$\alpha$ images
respectively at the same scale with north up and east to the left.  The
small crosses on the images denote the position of the nucleus but do
not indicate the extent of the Airy ring residuals.  Fig.  1c is again
Paschen $\alpha$ at a lower stretch with the radio contours of
\citet{gal96} superimposed. The radio source S1 is centered on the
nucleus. The expanded image scale is given by the tick marks on the
radio contours which are in units of arc seconds.  Fig. 1d shows the
radio image in black superimposed on the 12.5 $\micron$ contours with
the scale given in the axes. This image was taken directly from the
electronic version of \citet{bk00}.}

\label{fig-radpa}

\end{figure}

The black crosses in Fig.~\ref{fig-radpa}a and b indicates the location
of the nucleus in the Paschen $\alpha$ and H $\alpha$ emission images.
Fig.~\ref{fig-radpa}c shows the Paschen $\alpha$ image at a different
intensity stretch and spatial scale with the radio contours from
\citet{gal96} superimposed.  The units on the radio contour are in arc
seconds. The radio source S1 has been placed at the location of the
infrared point source as discussed in section~\ref{sec-nuc}.
Fig.~\ref{fig-radpa}d shows the radio emission image along with the
12.5 $\micron$ mid IR emission contours from \citet{bk00}.  The
emission region near the nucleus is best observed in H$\alpha$
(Fig.~\ref{fig-radpa} b) which is not affected by strong nuclear
emission.

After the bend in the radio jet the most intense hydrogen emission is
mainly to the north of the radio jet.  It is generally believed that
the bend in the jet is due to an interaction with a dense cloud.  The
emission region to the north may be the dense cloud which is being
ionized on its front face by the nuclear source. Note that the jet
points directly at an emission spot to the northeast that is seen both
in the H$\alpha$ and the Paschen $\alpha$ images.  The spot appears
faintly just above the upper left corner of the radio contours bounding
box in Fig.~\ref{fig-radpa}c.  The wispy arms of emission further to
the northeast have a similar appearance in both the H$\alpha$ and
Paschen $\alpha$ images.

\subsection{H Emission and Mid IR Emission}

Comparison of the morphology of the Paschen $\alpha$ emission and the
mid IR 12.5 $\micron$ emission (Fig.~\ref{fig-radpa}c and d) shows
remarkable similarity.  \citet{bk00} also placed their peak emission at
the location of S1.  They find the mid IR emission region falls north
of the bend in the radio jet in a feature they term the tongue.  Their
mid IR emission continues further south than the Paschen $\alpha$
emission, probably because the southern ionization cone penetrates into
the plane of the sky and the higher extinction attenuates the Paschen
$\alpha$ emission more than the mid IR emission.  A logical explanation
for the coincidence of emission is dust entrained into the HII region
ionized by the central source.  This is similar to Galactic HII regions
where the strengths of the mid IR emission and the hydrogen emission
lines are well correlated.  The lack of a true point source at the
nucleus in the mid IR observations is probably due to warm dust
extending further from the source than the extremely hot dust emission
that dominates the 2 $\micron$ emission.

\subsection{[SiVI] and [FeII] Emission Region}

\begin{figure}

\scalebox{.7}{\includegraphics{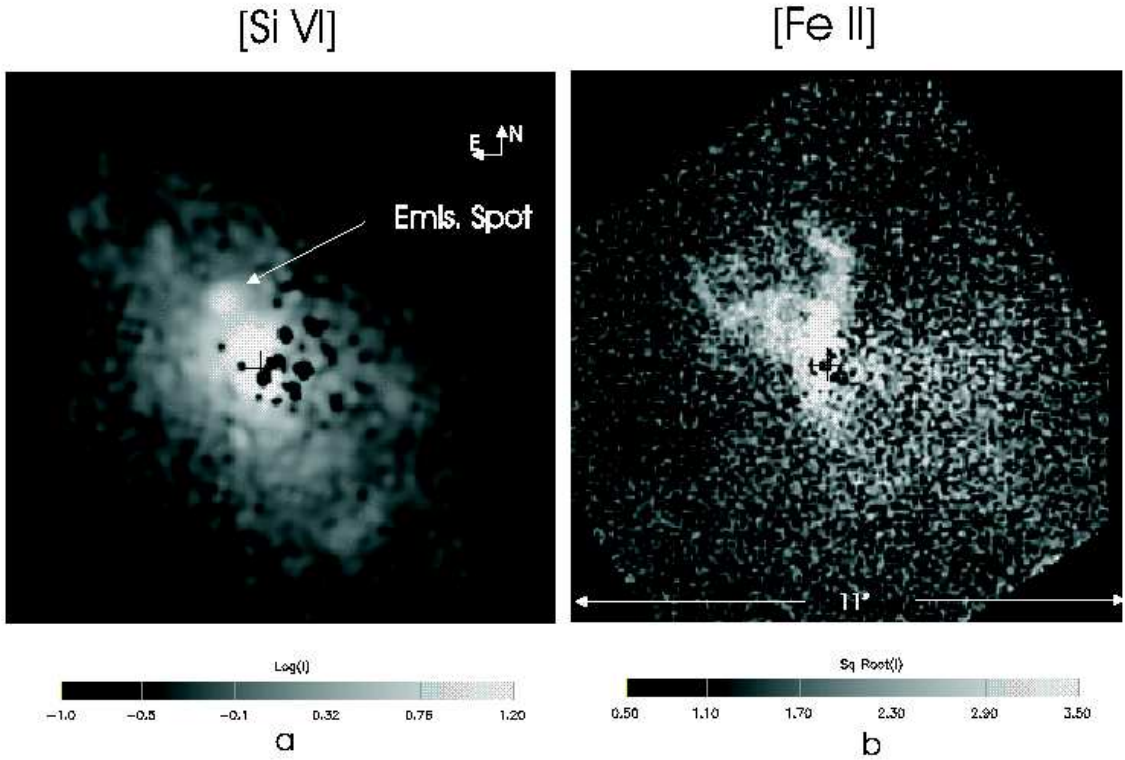}}

\caption{Images of the central region of NGC 1068 in [SiVI] and
[FeII]. The  [Si VI] image has a log stretch while the [FeII] image
has a square root stretch. As in Fig.~\ref{fig-radpa} north is up and
east is left. Although the [FeII] image was taken with camera 1 and
[SiVI] with camera 3, they have both been scaled to the camera 2 scale
for direct comparison with Fig.~\ref{fig-radpa}.}

\label{fig-si6}

\end{figure}

Fig.~\ref{fig-si6}a and b show the forbidden [SiVI] coronal emission
and [FeII] emission images  at the same spatial scale and location.
The [SiVI] image was taken by NICMOS camera 3 with $0.2 \arcsec$ pixels
as opposed to the $0.075 \arcsec$ pixel camera 2 Paschen $\alpha$
image. Within the differences of resolution, the [SiVI] and hydrogen
images are the same with strong emission near the nucleus and an
emission spot, cloud G of \citet{ev91}, along the direction of the
radio jet.  The emission spot in the Si image, however, is much
stronger relative to the total Si emission than its counterpart in
Paschen $\alpha$ emission.  The wispy extended hydrogen emission arms
are only faintly seen in the Si image even at a very high stretch. The
[SiVI] emission line strength is much less than the Paschen $\alpha$
line and the pixels are larger, hence the residuals near the nucleus
are much stronger.  It should be emphasized that the Paschen $\alpha$
and [SiVI] emission spot discussed here is not at the location of the
"hot spot" discussed by \citet{krm00} which is located near the
nucleus.

The [FeII] emission image is bright in the wispy arms but does not show
the emission spot seen in the Si and hydrogen images.  The [FeII]
image was taken with the $0.043 \arcsec$ pixel NICMOS camera 1 and has
poorer signal to noise than the other images.

\section{INTERPRETATION} \label{int}

The radio emission is generally interpreted as synchrotron and thermal
bremsstrahlung emission from a jet bent by an encounter with material
outside of the nucleus.  The excess mid-IR, Paschen $\alpha$ and [SiVI]
emission to the north and west of the radio jet may represent the
material that is producing the bend.  If the geometry discussed below
is correct this material has a direct view of the nucleus and is
ionized by its radiation field. The H and [SiVI] emission regions
monitor the material that can see the central engine and hence are
photoionized.  The analysis of the coronal [Si] emission by
\citet{thm96} showed that the [Si] line ratios were not consistent with
a shock heated emission region but were consistent with ionization by a
hard radiation field with a spectral index between $\nu^{-1}$ and
$\nu^{-1.5}$.

The location of  the secondary peak of emission, or emission spot, in
both H and [SiVI] along the direction of the radio jet, but beyond the
region of strong radio emission, suggests that the jet has impacted
material outside of the nucleus at this point. Further evidence for
material blocking the jet is the reduction of H emission beyond the
emission spot except in thin wispy arms on either side of the emission
spot.  A similar mechanism was proposed earlier by \citet{ax98} for the
emission region 0.1 arc seconds south of the emission spot.

The [SiVI] emission spot contains only 1/8 of the flux of the total [SiVI] emission.  The results of \citet{thm96}, which
could not resolve the two regions is, therefore, still consistent with
photoionization for the majority of the emission. The
analysis by \citet{al2000} of low spatial resolution NGC 1068 spectra
from several sources favors a two component ionizing mechanism.
\citet{krm00} also find another source of ionizing radiation for
emission at distances greater than 100 pc from the nucleus in the
northeast direction, which is the location of the emission spot.
\citet{krm00b} attribute the blue shifted emission at this location as
due to fast shocks resulting from the interaction of the emission line
knots with the interstellar medium.  This may be true for some of the
other emission knots but the geometry of this particular emission spot
seems to point to the jet as the most likely source of the
interaction.  The lack of [Fe II] emission at the spot is most likely
due to Fe being predominantly in higher ionization states at that
location.  This is supported by the observation of \citet{pec97} that
cloud G is a maximum in the [FeVIII]/[OIII] intensity ratio.

\section{CONCLUSIONS}

The NICMOS high resolution infrared images of NGC 1068 are consistent
with the general picture of a black hole nucleus, surrounded by an
obscuring torus, with ejection of material via jets perpendicular to
the plane of the torus.  Interaction of the jet with material 0.5 arc
seconds north of the nucleus deviates the jet as seen in the radio
emission maps.  This material is illuminated by the nucleus and can be
seen in both hydrogen and Si emission.  The jet eventually impacts
other circumnuclear material creating an emission spot in both H and
Si.  The coincidence of the Paschen $\alpha$ emission and the mid IR
features is probably due to dust entrained in the HII region in a
manner similar to that observed in Galactic HII regions.

\acknowledgments
 This work is supported in part by NASA grant NAG 5-3042.  This letter
 is based on observations with the NASA/ESA Hubble Space Telescope,
 obtained at the Space Telescope Science Institute, which is operated
 by the Association of Universities for Research in Astronomy under NASA
 contract NAS5-26555.  We would also like to acknowledge the helpful
suggestions and comments of the anonymous referee.

\clearpage

\clearpage

\end{document}